\documentclass[preprint,showpacs,preprintnumbers,amsmath,amssymb
              ,prb,eqsecnum]{revtex4}

\usepackage{graphicx}
\usepackage{dcolumn}
\usepackage{bm}

\begin{document}

\preprint{APS/123-QED}

\title{Theory of resonant inelastic x-ray scattering \\
       at the K edge in La$_2$CuO$_4$  \\
       - Multiple scattering effects -}

\author{Jun-ichi Igarashi$^1$, Takuji Nomura$^2$, and Manabu Takahashi$^3$}
\affiliation{%
$^1$Faculty of Science, Ibaraki University, Mito, Ibaraki 310-8512, Japan\\
$^2$Synchrotron Radiation Research Center, Japan Atomic Energy 
Agency, Hyogo 679-5148, Japan\\
$^3$Faculty of Engineering, Gunma University, Kiryu, Gunma 376-8515, Japan}

\date{\today}

\begin{abstract}
We develop a theory of resonant inelastic x-ray scattering (RIXS)
at the $K$ edge in La$_2$CuO$_4$ on the basis of the Keldysh
Green's function formalism. In our previous analysis 
(Phys. Rev. B 71, 035110 (2005)), the scattering 
by the core-hole potential was treated within the Born approximation,
and a crude-model density of states was used for the $4p$ band.
We improve the analysis by taking account of the multiple scattering 
in Cu$3d$-O$2p$ bands and by using a realistic $4p$ DOS 
obtained from a band calculation.
The multiple scattering effect is evaluated with the use of the time 
representation developed by Nozi\`eres and De Dominicis. 
It is found that the multiple scattering effect makes the $K$-edge peak 
in the absorption coefficient shift to the lower energy region 
as a function of photon energy, that is, the photon energy required to excite
the $1s$ electron to the $K$-edge peak reduces.
It is also found that the multiple-scattering effect does not change 
the two-peak structure in the RIXS spectra but modifies slightly the shape 
as a function of energy loss. 
These findings suggests that the multiple scattering effect 
could mainly be included into a renormalization of the core-level energy and
partly justify the Born approximation, leading to a future application to
the RIXS in three-dimensional systems.
\end{abstract}

\maketitle

\section{\label{sect.1}Introduction}
Resonant inelastic x-ray scattering (RIXS), taking advantage of the high
intensity of synchrotron sources, has become a powerful tool to probe
charge excitations in solids.\cite{Kao96,Hill98,Hasan00,Kim02,Inami03,Kim04-1,
Suga05} The momentum dependence is detectable
in transition-metal compounds by using the $K$-edge resonance,
because the wavelength of photon is an order of crystal lattice spacing.
The spectra consist of several peaks as a function of energy loss,
and the peak positions sometimes move with changing momenta transferred
to the crystal.
This is quite different from the optical conductivity,\cite{Uchida91}
where the momentum transfer is limited to nearly zero.
Electron energy-loss spectroscopy can also detect the momentum dependence 
on charge excitations, but it suffers from strong multiple scattering effects.
\cite{Wang96}
Therefore, RIXS is quite valuable to investigate charge excitations,
which is comparable to the neutron inelastic scattering for detecting 
spin excitations.

In the RIXS process, the $1s$ core electron is prompted to an empty $4p$ state 
by absorbing the photon, then charge excitations are created in order to screen 
the core-hole potential. Finally the photo-excited $4p$ electron is
recombined with the core hole by emitting the photon.
Charge excitations are left with energy and momentum transferred from the 
photon at the end. Most theoretical studies on the momentum dependence of
the RIXS spectra in cuprates have been based on the numerical diagonalization 
method for small clusters. The $4p$ band was replaced by a single level there.
\cite{Tsutsui99,Tsutsui03,Okada06} 
A single-band Hubbard model has been sometimes used with replacing the 
charge-transfer band (so-called ``Zhang-Rice" band) by the lower Hubbard band
in two-dimensional cases.\cite{Tsutsui99,Tsutsui03}
It is not clear whether this replacement is appropriate, 
since the low energy peak is assigned to
an excitation from the charge-transfer band to the upper Hubbard band.
The single-band Hubbard model cannot describe another high-energy peak, either.

By contrast, in our previous papers,\cite{Nomura04,Nomura05} 
we have formulated the RIXS spectra on the multiband tight-binding model
by adapting the resonant Raman theory developed by Nozi\`eres and Abrahams.
\cite{Noziere74} 
The formalism was based on the Hartree-Fock approximation (HFA) 
to describe the electronic states in the antiferromagnetic (AF) phase 
and the random-phase approximation (RPA) to the propagation of electron-hole 
pairs. The HFA is known to provide a good starting point for undoped materials.
The calculation reinforced by the RPA correction has reproduced well 
the experimental RIXS spectra as a function of energy loss 
and the dependence on momentum in La$_2$CuO$_4$.\cite{Nomura05}
The same formula has been applied to doped systems.\cite{Markiewicz06}
Thus, the formula seems promising to analyze the RIXS spectra in
more complicated systems in three dimensions. But it should be noted
that we have used a crude model for the $4p$ band
and have treated the core-hole potential in the intermediate state 
by the Born approximation with neglecting the multiple scattering effect. 
Since the core-hole potential is not weak, higher-order effects might be 
important. Therefore, it is desirable to check and improve the approximations 
made before going to further applications. 
One purpose of this paper is to address this issue by using a realistic 
$4p$ density of states obtained from the band calculation and 
by evaluating the higher-order effects.
Another purpose of this paper is to give a detailed derivation of
the formula of RIXS spectra, because details have been sketchy
in our previous papers.\cite{Nomura04,Nomura05}
We present the formula with an emphasis of the time representation.

We invent a numerical method to treat the multiple scattering by the core-hole 
potential along the line of Nozi\`eres and De Dominicis,\cite{Noziere69} 
deriving the Dyson equation to the one-electron Green's function with 
the core-hole potential working in a finite time interval. The point is that,
different from metallic systems, we can numerically solve the Dyson equation
for insulating systems, because the equation has no singular term.
On the basis of the solution, we evaluate the multiple scattering effects
on the RIXS spectra as well as the absorption coefficient.
This type of analysis has not been attempted before.
In the absorption coefficient, it is found that the $K$-edge peak 
moves to the lower energy region due to screening, 
and that the intensity at a higher
energy region is a little enhanced as an antiscreening effect.
When the incident photon energy is tuned at the $K$-edge peak,
the RIXS spectra have peaks at around 2 and 5 eV
as a function of energy loss.
The 2eV-peak corresponds to the excitation of electron 
from the charge transfer band to the upper Hubbard band.
It is found that the two-peak structure does not alter but the shape is 
slightly modified by the multiple scattering effect; 
the 2 eV peak is enhanced while the 5 eV peak is suppressed. 
The enhancement of the 2 eV peak is reasonable, because
the core-hole potential is most likely to be screened by the excitation
of electron from the charge-transfer band to the upper Hubbard band
in the intermediate state when the incident photon energy is tuned
at the $K$ edge.  However, the multiple scattering effect is limited 
within a small correction in the RIXS spectra. 
This may partly justify the Born approximation 
after a renormalization of the core-level energy.

The present paper is organized as follows.
In Sec. \ref{sect.2}, we introduce a model and formulate the RIXS spectra.
In Sec. \ref{sect.3}, the electronic structure is calculated on the
$d$-$p$ model within the HFA in the AF phase of La$_2$CuO$_4$.
The RIXS spectra are calculated within the Born approximation 
in comparison with the experiment in Sec. \ref{sect.4}.
The multiple scattering effect is evaluated
on the absorption and the RIXS spectra in Sec. \ref{sect.5}.
Section \ref{sect.6} is devoted to the concluding remarks.

\section{\label{sect.2}Formulation for RIXS spectra}

\subsection{Description of model}

We start by the expression of the Hamiltonian of photon,
\begin{equation}
 H_{ph}= \sum_{{\bf q}\alpha} \omega_{\bf q}
   c_{{\bf q}\alpha}^\dagger c_{{\bf q}\alpha},
\end{equation}
where operator $c_{{\bf q}\alpha}$ represents the annihilation operator of 
the photon with momentum ${\bf q}$, polarization $\alpha$.
For the interaction between photon and matter, we consider the dipole 
transition at the $K$ edge, where the $1s$ core electron is excited 
to the $4p$ band with absorbing photon and the reverse process takes place. 
This process may be described by
\begin{equation}
 H_x =w\sum_{{\bf q}\alpha}\frac{1}{\sqrt{2\omega_{\bf q}}}
      \sum_{j\eta\sigma}e^{(\alpha)}_{\eta}
       {p'}_{j\eta\sigma}^{\dagger}s_{j\sigma}
       c_{{\bf q}\alpha}{\rm e}^{i{\bf q}\cdot{\bf r}_j} + {\rm H. c.},
\label{eq.dipole}
\end{equation}
where $e^{(\alpha)}_{\mu}$ represents the $\eta$th component 
($\eta=x,y,z$) of two kinds of polarization vectors ($\alpha=1,2$) of photon.
Since the $1s$ state is so localized that the $1s\to 4p$ dipole transition 
matrix element is well approximated as a constant $w$.
Annihilation operators $p'_{j\eta\sigma}$ and $s_{j\sigma}$ are 
for states $4p_{\eta}$ and state $1s$ at Cu site $j$, respectively.

The Hamiltonians for the core electron and the $4p$ electron are given by
\begin{eqnarray}
 H_{1s} &=& \epsilon_{1s}\sum_{j\sigma}s_{j\sigma}^{\dagger}s_{j\sigma},\\
 H_{4p} &=& \sum_{{\bf k}\eta\sigma}\epsilon_{4p}^{\eta}({\bf k})
             {p'}_{{\bf k}\eta\sigma}^{\dagger}p'_{{\bf k}\eta\sigma}.
\end{eqnarray}
The photo-created $1s$ core hole induces charge excitations 
through the attractive core-hole potential, which may be described by
\begin{equation}
 H_{1s-3d} = V \sum_{j\sigma\sigma'} d_{j\sigma}^{\dagger}d_{j\sigma}
               s_{j\sigma'}^{\dagger}s_{j\sigma'}.
\end{equation}
Here $V$ may be $5-10$ eV in La$_2$CuO$_4$.
We neglect the interaction between the core-hole and the $4p$ electron,
since the $4p$ states are well extended in space with the bandwidth
as large as $\sim 20$ eV.
The excited $4p$ electron is finally recombined with the 
core-hole with emitting photon. In the end, charge excitations remain 
with receiving a momentum and an energy from scattering photons.

\begin{figure}
\includegraphics[width=8.0cm]{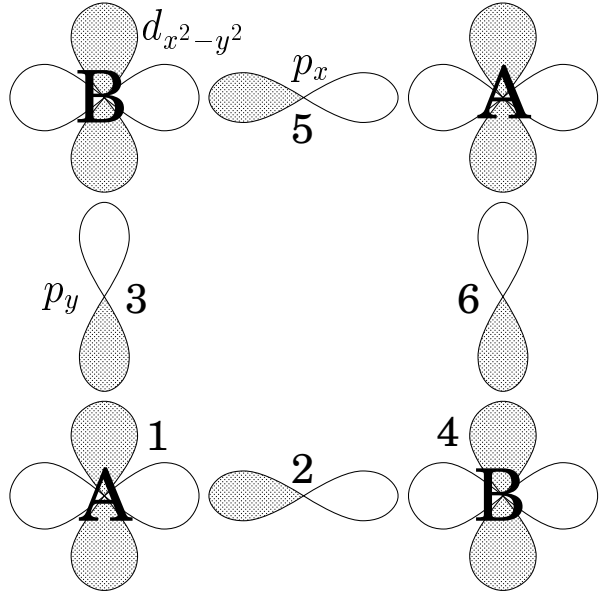}%
\caption{\label{fig.orbitals}
Schematic view of the unit cell in the antiferromagnetic phase. 
Six orbitals (numbered 1-6) are considered. There are two nonequivalent sites
of Cu shown as A and B. The gray parts in orbitals indicate that their 
wave functions take negative values.
}
\end{figure}

In La$_2$CuO$_4$, Cu and O atoms form a two-dimensional network shown in
Fig.~\ref{fig.orbitals}. There is nominally one hole per Cu atom.
To describe this situation, we consider only the $x^2-y^2$ orbital at the Cu 
site, which hybridizes the $\sigma$-bonding $2p$ orbitals at O sites (``$d$-$p$"
model). The corresponding Hamiltonian may be expressed as
\begin{eqnarray}
 H_{dp} &=& \epsilon_d \sum_{j\sigma}d_{j\sigma}^{\dagger}d_{j\sigma}
     + \epsilon_p \sum_{\ell\sigma}p_{\ell\sigma}^{\dagger}
                     p_{\ell\sigma}\nonumber \\
 &+& \sum_{j\ell\sigma} t_{j\ell}d_{j\sigma}^{\dagger}p_{\ell\sigma}+{\rm H.c.}
    +  \sum_{\ell\ell'\sigma}t_{\ell\ell'}
    p_{\ell\sigma}^{\dagger}p_{\ell'\sigma}
   \nonumber\\
   &+& U_{d}\sum_j d_{j\uparrow}^{\dagger}d_{j\uparrow}
               d_{j\downarrow}^{\dagger}d_{j\downarrow},
\end{eqnarray}
where $d_{j\sigma}$ is the annihilation operator for the $x^2-y^2$ 
orbital with spin $\sigma$ at Cu site $j$,
and $p_{\ell\sigma}$ is the annihilation operator of the $2p$ orbital 
with spin $\sigma$ at site $\ell$. 
For La$_2$CuO$_4$, the transfer energies between $3d$ and $2p$ orbitals $t_d$ 
and that between $2p$ orbitals $t_p$ are estimated from the local density
approximation (LDA) calculations, that is, $t_d=1.3$ eV, $t_p=0.65$ eV.
\cite{Hybertsen89}
We assume $U_d=11$ eV, and the $3d$ level energy relative to the $2p$ level
$\epsilon_d+\frac{1}{2}U_{d}n_d-\epsilon_p=-0.7$ eV with $n_d=\sum_{\sigma}
\langle d_{j\sigma}^{\dagger}d_{j\sigma}\rangle$.
These values are set the same as those in our previous paper.\cite{Nomura05}

\subsection{Keldysh-Schwinger formalism}

\begin{figure}
\includegraphics[width=8.0cm]{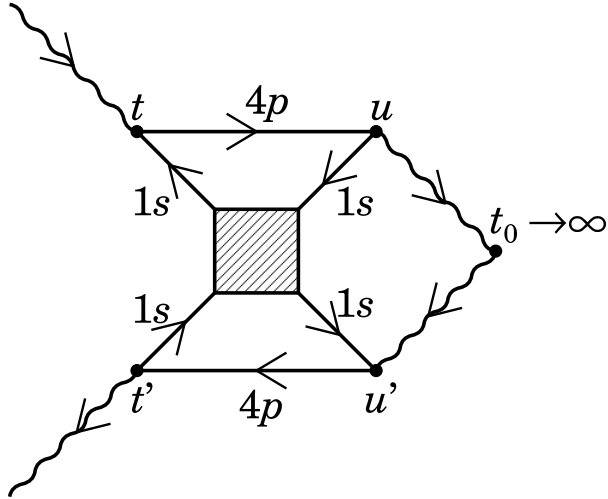}%
\caption{\label{fig.diagram2}
Expansion of the S matrix. The wavy lines represent photon Green's functions.
The solid lines with 4p and 1s represent Green's functions
of the $4p$ electron and the $1s$ core hole, respectively.
}
\end{figure}

Following Nozi\`eres and Abraham,\cite{Noziere74} 
we use the Keldysh-Schwinger formalism 
to the RIXS spectra. We prepare the initial state that 
one photon exists with ${\bf q}_i$, $\alpha_i$ in addition to 
a material in the ground state, which may be expressed as
\begin{equation}
  |\Phi_i\rangle = c_{{\bf q}_i\alpha_i}^{\dagger} |0\rangle , 
\end{equation}
where $|0\rangle$ is the ground state of the matter with no photon.
Let $H\equiv H_{ph}+H_{dp}+H_{1s}+H_{4p}$ be the unperturbed Hamiltonian 
of the system and $H_x$ be the perturbation. Then the $S$ matrix is given by
\begin{equation}
 U(t,-\infty) = T\exp\left\{-i\int_{-\infty}^t H_x(t'){\rm d}t'\right\},
\end{equation}
with $H_x(t)=\exp(iHt)H_x\exp(-iHt)$.
The probability of finding a photon with momentum ${\bf q}_f$, polarization
$\alpha_f$ at time $t_0$ is given by
\begin{equation}
 P_{{\bf q}_f\alpha_f;{\bf q}_i\alpha_i}(t_0)
  =\langle \Phi_i| U(-\infty,t_0)c_{{\bf q}_f\alpha_f}^\dagger
              c_{{\bf q}_f\alpha_f}U(t_0,-\infty) |\Phi_i\rangle .
\label{eq.prob}
\end{equation}
Expanding the S-matrix to second order in $H_x$, 
\begin{eqnarray}
 U(t,-\infty) &=& 1 + (-i)\int_{-\infty}^t H_x(t'){\rm d}t' \nonumber \\
              &+& \frac{(-i)^2}{2}\int_{-\infty}^t \int_{-\infty}^t
	        T(H_x(t')H_x(t'')){\rm d}t'{\rm d}t'',
\end{eqnarray}
we insert this into Eq.~(\ref{eq.prob}).
Figure \ref{fig.diagram2} shows a schematic representation of the expansion,
where the wavy lines indicate photons, which carry energy and momentum,
and the solid lines with ``4p" and ``1s" represent the Green's functions
of the $4p$ electron and the $1s$ core hole, respectively.
The upper and lower halves of the graph correspond to the so called
``outward" and ``backward" time legs, respectively.
By factoring out the dependence on the photon frequencies, we obtain
\begin{equation}
 P_{q_f\alpha_f;q_i\alpha_i}(t_0)
  = \int_{-\infty}^{t_0}{\rm d}u\int_{-\infty}^{u}{\rm d}t
    \int_{-\infty}^{t_0}{\rm d}u'\int_{-\infty}^{u'}{\rm d}t'
     S(t,u;t'u'){\rm e}^{i\omega_i(t'-t)}{\rm e}^{-i\omega_f(u'-u)}.
\end{equation} 
The transition probability per unit time with $t_{0}\to\infty$
is given by fixing one time, for instance, $u=0$,
\begin{equation}
 W(q_f\alpha_f;q_i\alpha_i)
  = \int_{-\infty}^{0}{\rm d}t
    \int_{-\infty}^{\infty}{\rm d}u'\int_{-\infty}^{u'}{\rm d}t'
     S(t,0;t'u'){\rm e}^{i\omega_i(t'-t)}{\rm e}^{-i\omega_fu'}.
\end{equation} 

\section{\label{sect.3}Hartree-Fock Approximation}

The undoped cuprates show the AF long-range order 
at $T=0$. It is known that the HFA works rather well in the AF phase
even for the case of large Coulomb interaction between $3d$ electrons.
Since the following analysis of RIXS is based on the HFA, we summarize
the electronic structure within the HFA in this section.
The unit cell in the AF phase contains six orbitals for each spin,
as shown in Fig.~\ref{fig.orbitals}.
Introducing the Fourier transform in the magnetic Brillouin zone (BZ),
we define the single-particle Green's function for six orbitals 
in a matrix form,
\begin{equation}
 [\hat G^{\sigma}({\bf k},\omega)]_{\mu\nu}
  = -i\int \langle T(A_{{\bf k}\mu\sigma}(t)
  A_{{\bf k}\nu\sigma}^{\dagger}(0))\rangle{\rm e}^{i\omega t}{\rm d}t,
\end{equation}
where $A_{{\bf k}\mu\sigma}$ represents the annihilation operator of the
electron with the orbital at site $\mu$ in the unit cell.
For example, at Cu sites ($\mu=1$ and $4$), it is given by
\begin{equation}
  A_{{\bf k}\mu\sigma}\equiv d_{{\bf k}\sigma}
      =\sqrt{\frac{2}{N}}\sum_i d_{i\sigma} {\rm e}^{i{\bf kr}_i}, 
\label{eq.m_rep}
\end{equation}
with $i$ running over the sites ``$\mu$" of $N/2$ unit cells.

\begin{figure}
\includegraphics[width=8.0cm]{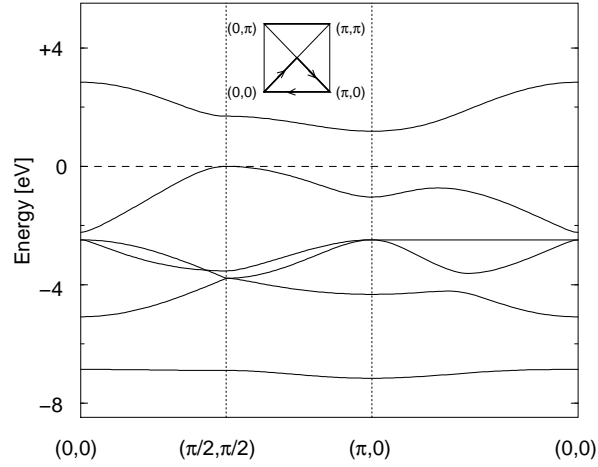}%
\caption{\label{fig.dispersion}
Dispersion relation of Cu$3d$-O$2p$ bands along symmetry lines
in the Brillouin zone. 
}
\end{figure}

Applying the equation-of-motion method to the Green's functions
we obtain the relation,
\begin{equation}
 (\omega\hat I-\hat J^{\sigma}({\bf k}))\hat G^\sigma({\bf k},\omega)
   = \hat I,
\end{equation}
where $\hat I$ is the unit matrix, and $\hat J^{\sigma}({\bf k})$ is 
given by
\begin{equation}
 \left(\begin{array}{cccccc}
  \epsilon_d+U_d n_{-\sigma}^A & t_{d}{\rm e}^{i\frac{k_x}{2}} 
  & -t_{d}{\rm e}^{i\frac{k_y}{2}} & 0 & -t_{d}{\rm e}^{-i\frac{k_x}{2}}
  & t_{d}{\rm e}^{-i\frac{k_y}{2}} \\
  & \epsilon_p &-2t_{p}\cos(\frac{k_x}{2}-\frac{k_y}{2}) 
  &-t_{d}{\rm e}^{i\frac{k_y}{2}} & 0 
  & 2t_{p}\cos(\frac{k_x}{2}+\frac{k_y}{2})\\
  & & \epsilon_p & t_{d}{\rm e}^{i\frac{k_y}{2}} 
  & 2t_{p}\cos(\frac{k_x}{2}+\frac{k_y}{2}) & 0 \\
  & & & \epsilon_d+U_d n_{-\sigma}^B & t_{d}{\rm e}^{i\frac{k_x}{2}} 
  & -t_{d}{\rm e}^{i\frac{k_y}{2}} \\
  & & & & \epsilon_p &-2t_{p}\cos(\frac{k_x}{2}-\frac{k_y}{2})\\
  & & & & & \epsilon_p 
	\end{array}\right) ,
\label{eq.hf_matrix}
\end{equation}
with 
\begin{equation}
 n_{\sigma}^{A}=\frac{2}{N}\sum_{\bf k}
      \int [\hat G^{\sigma}({\bf k},\omega)]_{11}
                     {\rm e}^{i\omega 0^+}\frac{{\rm d}\omega}{2\pi}, \quad
 n_{\sigma}^{B}=\frac{2}{N}\sum_{\bf k}
      \int [\hat G^{\sigma}({\bf k},\omega)]_{44}
                     {\rm e}^{i\omega 0^+}\frac{{\rm d}\omega}{2\pi}.
\end{equation}
Since $n_{\sigma}^{B}=n_{-\sigma}^{A}$, we may put
\begin{equation}
 n_{\uparrow\downarrow}^{A}=\frac{1}{2}(n\pm m),\quad 
 n_{\uparrow\downarrow}^{B}=\frac{1}{2}(n\mp m) . 
\end{equation}
Lower triangle components are the Hermitian conjugates to
the upper triangle components, which are omitted in Eq.~(\ref{eq.hf_matrix}).
The $\hat J^{\sigma}({\bf k})$ is diagonalized by an unitary matrix 
$\hat U^{\sigma}({\bf k})$, that is, $[\hat U^{-1}\hat J\hat U]_{jj'}
 = E_j({\bf k})\delta_{jj'}$.
Then the Green's function is expressed as
\begin{equation}
 \hat G^{\sigma}({\bf k},\omega) = U^{\sigma}({\bf k})
 \hat D({\bf k},\omega) U^{\sigma}({\bf k})^{-1},
\end{equation}
with 
\begin{equation}
 [\hat D({\bf k},\omega)]_{jj'}=
   \frac{1}{\omega-E_j({\bf k})\pm i\delta}
      \delta_{jj'}.
\end{equation}
Figure \ref{fig.dispersion} shows the dispersion relation $E_j({\bf k})$
as a function of ${\bf k}$ along symmetry lines. The conduction band
and the top of the valence band may be called the ``upper Hubbard" band and
the ``charge-transfer" band, respectively.

\section{\label{sect.4}Born approximation for RIXS spectra}

\begin{figure}
\includegraphics[width=8.0cm]{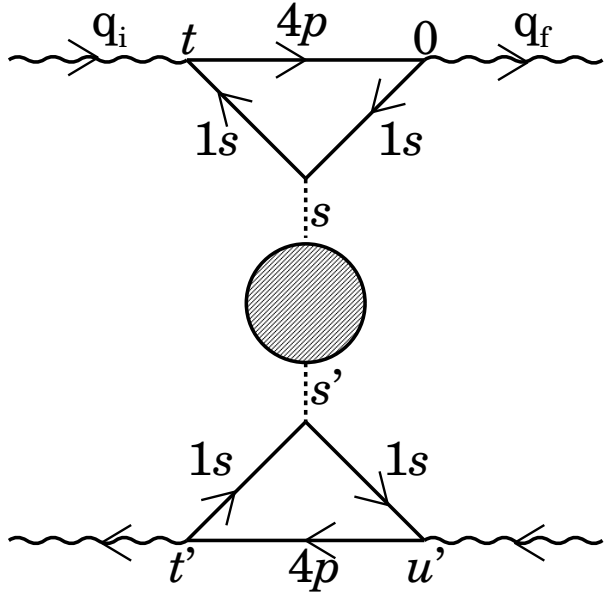}%
\caption{\label{fig.diagram1}
Diagrams for the RIXS intensity within the Born approximation. 
The dotted lines represent the core-hole potential.
The shaded part represents the Keldysh-type Green's function, which connects 
the outward time leg on the top half and the backward time leg 
on the bottom half.
}
\end{figure}

We consider the process that an electron-hole pair is singly created
to screen the core-hole potential. 
The multiple-scattering effect beyond the Born approximation is neglected.
Figure \ref{fig.diagram1} displays the corresponding diagrams,
where the shaded part represents the Keldysh-type Green's function,
\begin{eqnarray}
 Y_{\mu'\sigma',\mu\sigma}^{+-}({\bf q},s'-s) 
 &=& \int Y_{\mu'\sigma',\mu\sigma}^{+-}({\bf q},\omega) 
     {\rm e}^{-i\omega(s'-s)}\frac{{\rm d}\omega}{2\pi} \\
 &=& \langle \rho_{{\bf q}\mu'\sigma'}(s')
 \rho_{{\bf -q}\mu\sigma}(s)\rangle,
\end{eqnarray}
with
\begin{equation}
 \rho_{{\bf q}\mu\sigma}=\sqrt{\frac{2}{N}}\sum_{\bf k} 
     d^{\dagger}_{{\bf k+q}\mu\sigma}d_{{\bf k}\mu\sigma}.
\end{equation}
The momentum conservation requires the relation ${\bf q}={\bf q}_i-{\bf q}_f$,
and ${\bf k}$ runs over the magnetic first BZ. The superscripts $+$ and
$-$ stand for the backward and outward time legs, respectively.\cite{Landau62}
In the lowest order, it is given by
\begin{equation}
 Y_{\mu'\sigma',\mu\sigma}^{+-(0)}({\bf q},s'-s) 
 = \frac{2}{N}\sum_{\bf k}
 \langle d_{{\bf k+q}\mu'\sigma'}(s')d_{{\bf k+q}\mu\sigma}^{\dagger}(s)\rangle
 \langle d_{{\bf k}\mu'\sigma'}^{\dagger}(s')d_{{\bf k}\mu\sigma}(s)\rangle,
\end{equation}
and thereby 
\begin{eqnarray}
 Y_{\mu'\sigma',\mu\sigma}^{+-(0)}({\bf q},\omega) 
 &=& \delta_{\sigma\sigma'}\sum_{\bf k}\sum_{j,j'}
    \delta(\omega-E_j([{\bf k+q}])+E_j({\bf k}))
    [1-n_{j'}([{\bf k+q}])]n_j({\bf k}) \nonumber \\
  &\times& {\tilde U}^{\sigma}_{\mu'j'}([{\bf k+q}])
   {\tilde U}^{\sigma *}_{\mu j'}([{\bf k+q}]) 
   U^{\sigma}_{\mu j}({\bf k})U^{\sigma*}_{\mu'j}({\bf k}),
\label{eq.keldysh1}
\end{eqnarray}
where $j$ and $j'$ stand for energy eigenstates, and $[{\bf k+q}]$ is 
the reduced value of ${\bf k+q}$ into the magnetic first BZ by a reciprocal 
lattice vector ${\bf G}$, that is, ${\bf k+q}=[{\bf k+q}]+{\bf G}$. 
Since two Cu sites in the unit cell are involved to excite an electron-hole 
pair, phase factor ${\rm e}^{i{\bf Ga}}$ is attached to the wave function 
at the B site,
\begin{equation}
 {\tilde U}^{\sigma}_{\mu j}([{\bf k+q}])
    = {\rm e}^{i{\bf Ga}}U^{\sigma}_{\mu j}([{\bf k+q}])
    \quad {\rm for}\,\,\mu=4,
\end{equation}
where $\mu=4$ corresponds to the Cu B site, 
and ${\bf a}$ [$\equiv (a,0)$] represents
a position of the B site relative to the A site in the unit cell.

The product of Green's functions of the $4p$ electron and the core hole 
gives simply a factor 
$\exp[i(\epsilon_{4p}^{\eta}({\bf p})-\epsilon_{1s}-i\Gamma_{1s}-\omega_i)t]$
on the outward time leg and a factor 
$\exp[-i(\epsilon_{4p}^{\eta}({\bf p})-\epsilon_{1s}+i\Gamma_{1s}-\omega_i)
(t'-u')]$ on the backward time leg, 
where $\Gamma_{1s}$ is a lifetime broadening width of the $1s$ core hole. 
In a conventional calculation based on the Fourier transforms, 
this feature is hidden.
This property will be fully utilized in the study of the multiple scattering 
effect in the next section.
The Keldysh Green's function carries the time dependent factor 
${\rm e}^{i\omega s}$ to the outward time leg and ${\rm e}^{-i\omega s'}$
to the backward time leg. Note that the core-hole potential works only 
in intervals $[t,0]$ and $[t',u']$.
Integrating the time factor combined to the above product of Green's 
functions, with respect to $s$ and $t$ in the region of $t<s<0$, $-\infty<t<0$,
we obtain
\begin{eqnarray}
 L_B^{\eta}(\omega_i;\omega) &\equiv&
   V\int_{-\infty}^0{\rm d}t \sum_{\bf p}
   \exp[i(\epsilon_{4p}^{\eta}({\bf p})-\epsilon_{1s}-i\Gamma_{1s}-\omega_i)t]
  \int_{t}^0 {\rm d}s\,{\rm e}^{i\omega s} \nonumber \\
     &=& \int\frac{V\rho^{\eta}_{4p}(\epsilon){\rm d}\epsilon}
  {(\omega_i+\epsilon_{1s}+i\Gamma_{1s}-\epsilon)
           (\omega_i-\omega+\epsilon_{1s}+i\Gamma_{1s}-\epsilon)}.
\label{eq.born}
\end{eqnarray}
Here the sum over $4p$ states is replaced by the integration of 
the $4p$ DOS projected onto the $\eta$ ($=x,y,z$) symmetry,
$\rho_{4p}^{\eta}(\epsilon)$.
A similar factor has been derived in third-order perturbation theory
by Abbamonte et al.\cite{Abbamonte99}
The integration with respect to $s'$ and $t'$ in the backward time leg gives 
the term complex conjugate to Eq.~(\ref{eq.born}). 
The integration with respect to $u'$ gives the energy-conservation factor, 
which guarantees that $\omega$ in Eq.~(\ref{eq.born}) is the energy loss,
$\omega=\omega_i-\omega_f$.

\begin{figure}
\includegraphics[width=8.0cm]{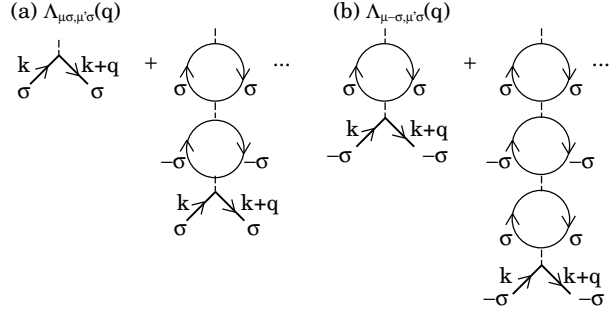}%
\caption{\label{fig.diagram4}
Vertex function within the RPA; (a) the channel of keeping spin, 
and (b) the channel of reversing spin.
Broken lines between bubbles 
represent the $3d$ Coulomb interaction,
which is effective only between $3d$ electrons with opposite spins.
}
\end{figure}

We evaluate the higher-order effect on 
$Y^{+-}_{\mu'\sigma',\mu\sigma}(q)$ [$q\equiv({\bf q},\omega)$]
within the RPA. Figure \ref{fig.diagram4} shows the corresponding diagram.
Collecting up the ladder diagrams, we obtain 
\begin{equation}
 Y^{+-}_{\mu'\sigma',\mu\sigma}(q)
  = \sum_{\mu_1\sigma_1\mu_2\sigma_2}
   \Lambda^{*}_{\mu_1\sigma_1,\mu'\sigma'}(q)
   Y^{+-(0)}_{\mu_1\sigma_1,\mu_2\sigma_2}(q)
   \Lambda_{\mu_2\sigma_2,\mu\sigma}(q),
\end{equation}
where
\begin{equation}
  \Lambda_{\mu_1\sigma_1,\mu_2\sigma_2}(q)
 = \left[ I-{\hat U}_d {\hat F}(q)\right]^{-1}_{\mu_1\sigma_1,\mu_2\sigma_2},
\end{equation}
with
\begin{eqnarray}
 \hat F(q) &=& \left( \begin{array}{cccc}
      F^{\uparrow}_{11}(q)   & 0 & F^{\uparrow}_{14}(q)   & 0 \\
      0 & F^{\downarrow}_{11}(q) & 0 & F^{\downarrow}_{14}(q) \\ 
      F^{\uparrow}_{41}(q)   & 0 & F^{\uparrow}_{44}(q)   & 0 \\
      0 & F^{\downarrow}_{41}(q) & 0 & F^{\downarrow}_{44}(q) 
         \end{array}\right), \\
 \hat U_d &=& \left( \begin{array}{cccc}
      0 & U_d & 0 & 0 \\
      U_d & 0 & 0 & 0 \\
      0 & 0 & 0 & U_d \\
      0 & 0 & U_d & 0
         \end{array}\right).
\end{eqnarray}
Here the polarization propagator $F^{\sigma}_{\mu\mu'}(q)$ is given by
\begin{eqnarray}
 F^{\sigma}_{\mu\mu'}(q) &=&
  -i\frac{2}{N}\sum_{\bf k}\int\frac{{\rm d}k_0}{2\pi}
    G^{\sigma}_{\mu\mu'}({\bf k},k_0)
    G^{\sigma}_{\mu'\mu}({\bf k+q},k_0+\omega) \nonumber \\ 
    &=&
    \frac{2}{N}\sum_{\bf k}
    U^{\sigma}_{\mu j}({\bf k})U^{\sigma *}_{\mu' j}({\bf k})
    {\tilde U}^{\sigma}_{\mu' j'}([{\bf k+q}]) 
    {\tilde U}^{\sigma *}_{\mu j'}([{\bf k+q}]) \nonumber \\
    &\times&\left[\frac{n_j({\bf k})[1-n_{j'}([\bf{k+q}])]}
    {\omega-E_{j'}([{\bf k+q}])+E_{j}({\bf k})+i\delta}
    -\frac{n_{j'}([{\bf k+q}])[1-n_j({\bf k})]}
    {\omega-E_{j'}([{\bf k+q}])+E_{j}({\bf k})-i\delta}\right].
\end{eqnarray}

Combining all these relations together, we finally obtain an expression of
the RIXS intensity,
\begin{eqnarray}
 W(q_i,\alpha_i;q_f,\alpha_f) 
 &=& 2\pi\frac{|w|^4}{4\omega_i\omega_f} 
 \sum_{{\bf k}jj'}\sum_{\mu\sigma\mu'\sigma'}
 \sum_{\mu_1\mu_2\bar{\sigma}}
 \delta(\omega+E_j({\bf k})-E_{j'}([{\bf k+q}]))n_j({\bf k})
 [1-n_{j'}([{\bf k+q}])] \nonumber\\
 &\times& \Lambda^{*}_{\mu_1\bar{\sigma},\mu'\sigma'}(q)
      U^{\bar{\sigma'}}_{\mu_2 j}({\bf k}) U^{\bar{\sigma}*}_{\mu_1 j}({\bf k})
     \tilde{U}^{\bar{\sigma}}_{\mu_1 j}([{\bf k+q}])
     \tilde{U}^{\bar{\sigma}*}_{\mu_2 j}([{\bf k+q}])
     \Lambda_{\mu_2\bar{\sigma},\mu\sigma}(q)
 \nonumber\\
 &\times&\left|\sum_{\eta}e_{\eta}^{(\alpha)}
  L^{\eta}_B(\omega_i;\omega)
 e_{\eta}^{(\alpha')} \right|^2,
\label{eq.general}
\end{eqnarray}
where the incident photon has the momentum and energy
$q_i=({\bf q}_i,\omega_i)$, polarization $\bf{\rm e}^{(\alpha_i)}$,
and the scattered photon has the momentum and energy
$q_f=({\bf q}_f,\omega_f)$, polarization $\bf{\rm e}^{(\alpha_f)}$.
The transfer of momentum and energy is denoted as
${\bf q}={\bf q}_i-{\bf q}_f$, $\omega=\omega_i-\omega_f$.

\begin{figure}
\includegraphics[width=8.0cm]{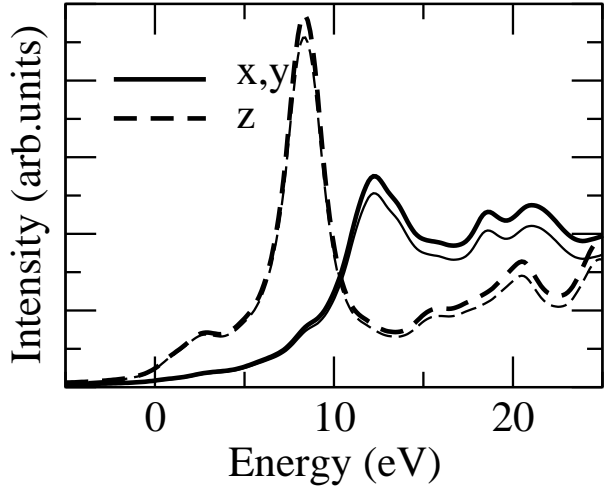}%
\caption{\label{fig.4pdos}
Density of states for the $4p$ band convoluted with the Lorentzian function
with HMFW $2\Gamma_{1s}=1.6$ eV.
The origin of energy corresponds to the bottom of the $4p$ band.
Thin lines represent the absorption coefficient with taking account of
the dipole matrix element evaluated by the band calculation.
}
\end{figure}

We have already reported the same formula without a detailed derivation 
in Ref. \onlinecite{Nomura05}. 
In that study, we calculated the RIXS spectra of La$_2$CuO$_4$
by assuming the two-dimensional cosine dispersion for the $4p$ band
and tuning the incident photon energy to excite the core electron to the peak
of the $4p$ DOS. We improve the treatment by using the
realistic $4p$ DOS calculated with the LDA.
Although the LDA fails to predict the AF insulating ground state, 
the $4p$ band is expected to be properly described because of weak 
correlation in wide bands.
Figure \ref{fig.4pdos} displays the $4p$ DOS convoluted
by the Lorentzian function with the half maximum full width (HMFW)
$2\Gamma_{1s}=1.6$ eV.
Reflecting the tetragonal structure, the $p_z$-symmetric DOS is found quite
different from the $p_{x}$- and $p_y$-symmetric DOS's.
Since the experiment by Kim {\it et al.} has been performed with the 
polarization along the $z$ axis, we concentrate our calculation to
the $p_z$-symmetric DOS.
Thin lines represent the absorption coefficient with taking account of
the $1s\to 4p$ dipole matrix element calculated by the band calculation
(the origin of energy corresponds to the excitation of the $1s$ electron
to the bottom of the $4p$ band).
The curves are very close to the $4p$ DOS, indicating that replacing
the dipole matrix element by a constant value in Eq.~(\ref{eq.dipole})
is a good approximation.

\begin{figure}
\includegraphics[width=8.0cm]{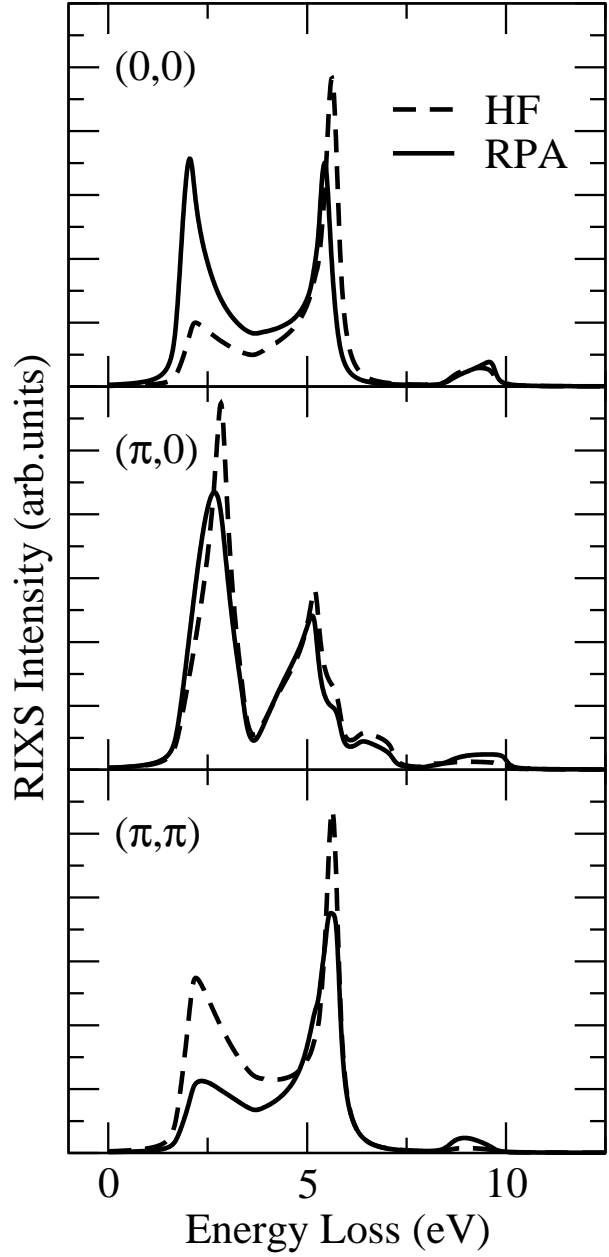}%
\caption{\label{fig.spec1}
RIXS spectra as a function of energy loss $\omega=\omega_i-\omega_f$,
within the Born approximation.
Momentum transfer is $(0,0)$, $(\pi,0)$, and $(\pi,\pi)$.
The incident photon has the $z$-polarization and the energy corresponding to
the excitation from the $1s$ core to the peak of the $p_z$ DOS.
The dotted and solid lines are the results of the HFA and RPA, respectively.
}
\end{figure}

We calculate the RIXS intensity from Eq.~(\ref{eq.general})
by tuning the incident photon energy at the peak of the $p_z$ DOS,
the $\delta$ function is replaced by the Lorentzian function with
HMFW$=0.2$ eV in order to take account of the instrumental resolution.
Figure \ref{fig.spec1} shows the calculated spectra as a function of
energy loss for momentum transfer $(0,0)$, $(\pi,0)$, and $(\pi,\pi)$.
It should be noted here that changing magnitude of $V$ does not change 
the spectral shape within the Born approximation;
as is evident from Eq.~(\ref{eq.general}), $W(q_i,\alpha_i;q_f,\alpha_f)/V^2$
is independent of $V$.
The present result using the realistic $4p$ DOS is nearly the same
as our previous one\cite{Nomura05} using a two-dimensional cosine dispersion 
for the $4p$ band. We obtain the spectra consisting of two peaks 
at about 2 and 5 eV as a function of energy loss,
whose shapes move with changing momenta in good agreement with the experiment.
The 2 eV peak is assigned to the excitation from the charge-transfer band 
to the upper Hubbard band.
The 2eV peak at $(0,0)$ is considerably enhanced, while that 
at $(\pi,\pi)$ is considerably suppressed by the vertex function of RPA 
in agreement with the experiment.
As already emphasized in our previous paper,\cite{Nomura05} 
the intersite correlation reinforced by the RPA is quite important 
for the quantitative understanding of the RIXS spectra.
The difference between the present and the previous results is that
the 2 eV peak is a little enhanced relative to
the 5 eV peak in the present result.

\section{\label{sect.5}Multiple scattering by the core-hole potential} 

Since the $1s$ core-hole potential is rather strong, 
many electron-hole pairs may be created in the Cu$3d$-O$2p$ bands
in order to screen the potential. This process may influence not only 
the absorption spectra but also the RIXS spectra.

It is known in metallic systems that the creation of an infinite number 
of electron-hole pairs leads to ``singular" behavior in the absorption spectra.
\cite{Mahan67,Noziere69}
In insulating systems, however, 
the number of electron-hole pairs is limited due to the finite energy gap,
and the singularity would not come out. Under this condition, 
we calculate the RIXS and absorption spectra on the basis of the time
representation developed by Nozi\`eres and De Dominicis.\cite{Noziere69}

\begin{figure}
\includegraphics[width=8.0cm]{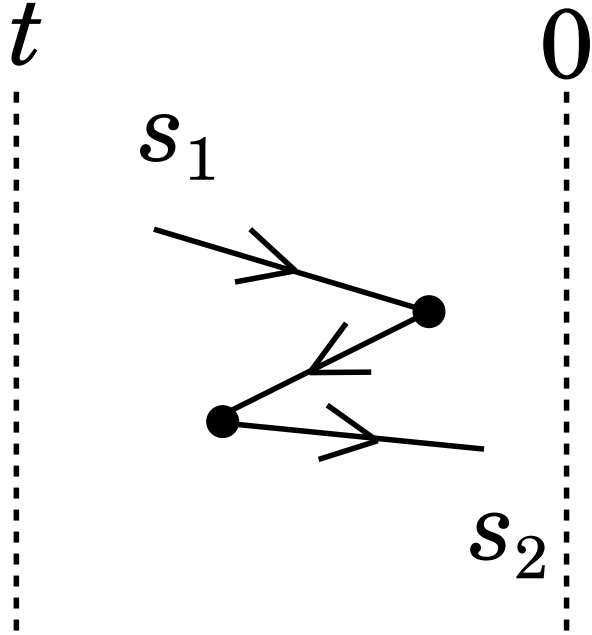}%
\caption{\label{fig.diagram_green}
Diagram for the one-electron Green's function $\varphi^{\sigma}(s_2,s_1)$.
Solid circles represent the core hole potential working on the electron.
}
\end{figure}

Since the potential is localized at Cu sites, 
it is convenient to introduce the one-electron Green's function 
for the $3d$ electron localized at the $\mu$th site ($\mu=1$ and $4$),
\begin{equation}
 \varphi_{\mu\sigma}(s_2,s_1) = -i
      \langle T(d_{\mu\sigma}(s_2)d_{\mu\sigma}^{\dagger}(s_1)\rangle.
\end{equation}
Here the core-hole potential works only between the time interval $[t,0]$, 
and $t < s_1, s_2 < 0$.
Figure \ref{fig.diagram_green} shows a diagram contributing 
to $\varphi_{\mu\sigma}(s_2,s_1)$.
Collecting similar diagrams, we obtain the Dyson equation 
\begin{equation}
 \varphi_{\mu\sigma}(s_2,s_1) = G_{\mu\sigma}^{(0)}(s_2-s_1) - 
  V\int_t^0 G_{\mu\sigma}^{(0)}(s_2-s_3) \varphi_{\mu\sigma}(s_3,s_1){\rm d}s_3,
\label{eq.dyson}
\end{equation}
where $G_{\mu\sigma}^{(0)}(s)$ represents the unperturbed local Green's 
function, which is given by
\begin{eqnarray}
  G_{\mu\sigma}^{(0)}(s) 
    &=& (-i)\frac{2}{N}\sum_{{\bf k}j}|U_{\mu j}^{\sigma}({\bf k})|^2
       (1-n_j({\bf k})){\rm e}^{-iE_j({\bf k})s},
       \quad s > 0, \\
    &=& i\frac{2}{N}\sum_{{\bf k}j}|U_{\mu j}^{\sigma}({\bf k})|^2
       n_j({\bf k}){\rm e}^{-iE_j({\bf k})s},
       \quad s < 0.
\end{eqnarray}

In metallic systems, $G_{\mu\sigma}^{(0)}(s)$ behaves as $1/s$ 
for $|s|>>D^{-1}$ with $D$ being the bandwidth. In this situation,
the Dyson equation becomes a singular integral equation, 
which needs special care.\cite{Noziere69} 
On the other hand, in an insulating system we are dealing with, 
the asymptotic term of $1/s$ would vanish for $s>>d^{-1}>D^{-1}$ with $d$ being
the band gap, owing to the presence of the energy gap. 
In addition, the solutions for large values of $|t|$ 
are less important in the absorption coefficient and the RIXS spectra,
since the finite core-hole life time reduces their contributions.
Therefore, we are allowed to solve the integral equation [Eq.~(\ref{eq.dyson})]
in the finite interval $[t,0]$ with setting the maximum value of $|t|$ 
by $12$ (eV)$^{-1}$. In the numerical solution, we divide the interval 
into small widths by $0.02$ (eV)$^{-1}$.
The reduction factor at $|t|=12$ (eV)$^{-1}$ due to the core-hole life-time 
width $\Gamma_{1s}=0.8$ eV is estimated as $\exp(-12\Gamma_{1s})=0.000067$, 
which seems sufficiently small.
Note that we need the one-electron Green's functions 
only for $s_2=t$ or $s_1=t$ in later use. 

The appropriate value of $V$ is not known and may be model dependent.
It may be better to take the value as a parameter. In the following,
we calculate the absorption and RIXS spectra for $V=6$ and $9$ eV,
in order to get some insight into the parameter dependence.

\subsection{Absorption coefficient}

\begin{figure}
\includegraphics[width=8.0cm]{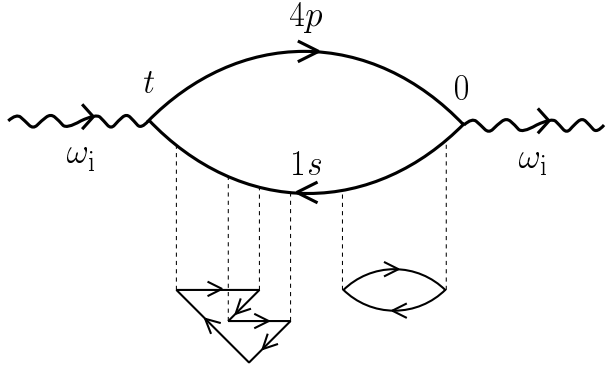}%
\caption{\label{fig.diagram_abs}
Diagram for the absorption coefficient at the Cu K-edge.
A photon enters with frequency $\omega$ and polarization $\eta$.
The solid lines with 4p and 1s represent the Green's function
of the $4p$ electron and $1s$ core hole, respectively.
Other solid lines represent the local Green's function.
The dotted lines denote the core-hole potential.
}
\end{figure}

Before going to the RIXS spectra, we first formulate the absorption spectra.
Let a photon have frequency $\omega_i$ and polarization direction $\eta$.
Then, the absorption coefficient is calculated from
\begin{equation}
 A_{\eta}(\omega_i) = \frac{1}{\pi}{\rm Re}\int_{-\infty}^0
      \langle H_x(0)H_x(t)\rangle{\rm d}t .
\end{equation}
Figure \ref{fig.diagram_abs} shows a corresponding diagram,
where the interaction between the core hole and the $4p$ electron is neglected.
To screen the core-hole potential, electron-hole pairs are created.
Without such bubble diagrams, the absorption coefficient is simply given by
\begin{eqnarray}
 A_{\eta}(\omega_i) &=& \frac{1}{\pi}\frac{|w|^2}{2\omega_i}
    {\rm Re} \sum_{\bf k}
    \int_{-\infty}^0 \exp\{i(\epsilon_{4p}^{\eta}({\bf k})-\epsilon_{1s}
                             -i\Gamma_{1s}-\omega_i)t\}
    {\rm d}t, \nonumber\\
           &=& \frac{|w|^2\Gamma_{1s}}{2\pi\omega_i}
	   \int\frac{\rho^{\eta}_{4p}(\epsilon)}
	   {(\omega_i-\epsilon+\epsilon_{1s})^2+\Gamma_{1s}^2}
		{\rm d}\epsilon.
\label{eq.abs_non}
\end{eqnarray}
This is nothing but the $4p$ DOS convoluted with the Lorentzian function
with HMFW $2\Gamma_{1s}$.

The electron-hole pairs are treated on the basis of the linked cluster 
expansion. The contribution is simply given by a factor ${\rm e}^{C(t)}$, 
where
\begin{equation}
 C(t) = -\frac{1}{2}V^2\sum_{\sigma}\int_{t}^0\int_{t}^0
        \varphi_{\mu\sigma}(s_2,s_1)G_{\mu\sigma}^{(0)}(s_1-s_2)
	{\rm d}s_1{\rm d}s_2.
\label{eq.linked}
\end{equation}
In the actual calculation of $C(t)$, we numerically evaluate 
${\rm d}C(t)/{\rm d}t$ from the relation
\begin{equation}
 \frac{{\rm d}C(t)}{{\rm d}t}
  = \frac{1}{2}V^2\sum_{\sigma}\int_t^0\left[
   \varphi_{\mu\sigma}(t,s)G_{\mu\sigma}^{(0)}(s-t)
   +\varphi_{\mu\sigma}(s,t)G_{\sigma}^{(0)}(t-s)
   \right]{\rm d}s,
\label{eq.linked_diff}
\end{equation}
and then integrate this quantity with respect to $t$,
instead of directly evaluating Eq.~(\ref{eq.linked}).
The product of Green's functions of the $4p$ electron and the core hole 
gives a factor 
$\exp[i(\epsilon_{4p}^{\eta}({\bf k})-\epsilon_{1s}-i\Gamma_{1s}-\omega_i)t]$, 
which is independent of times at which dotted lines are attached 
to the core-hole line in Fig.~\ref{fig.diagram_abs}.
Therefore, the absorption coefficient is given by
\begin{equation}
 A_{\eta}(\omega_i) = \frac{|w|^2}{2\pi\omega_i}
    \lim_{T\to -\infty}{\rm Re}\left[K^{\eta}(\omega_i;T)\right], 
\end{equation}
with
\begin{eqnarray}
    K^{\eta}(\omega_i;T)&=&\sum_{\bf k}\int_{T}^{0}
       \exp[i(\epsilon_{4p}^{\eta}({\bf k})-\epsilon_{1s}
              -i\Gamma_{1s}-\omega_i)t]{\rm e}^{C(t)} {\rm d}t \nonumber\\
      &=& \int\rho_{4p}^{\eta}(\epsilon){\rm d}\epsilon\int_{T}^{0}
       \exp[i(\epsilon-\epsilon_{1s}-i\Gamma_{1s}-\omega_i)t]
       {\rm e}^{C(t)} {\rm d}t.
\label{eq.Kfunc}
\end{eqnarray}

\begin{figure}
\includegraphics[width=8.0cm]{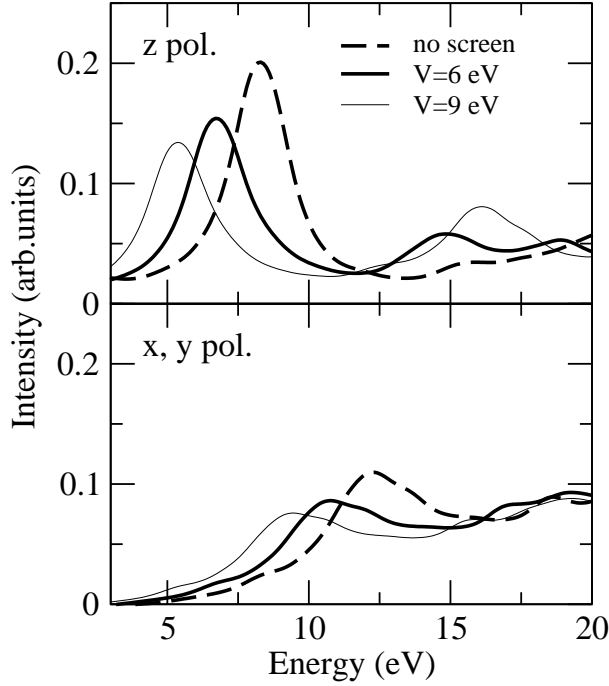}%
\caption{\label{fig.abs_abc}
Absorption coefficient for $z$, and $x,y$ polarizations.
The broken line represents the result without the screening.
The thick and thin solid lines represent the results with the screening by 
electron-hole pairs for $V=6$ and $9$ eV, respectively.
The origin of energy corresponds to the bare energy exciting the $1s$
electron to the bottom of the $4p$ band, that is, $\epsilon_{4p}^0
-\epsilon_{1s}$ with $\epsilon_{4p}^0$ being the energy of the $4p$ band
bottom.
}
\end{figure}

We integrate numerically Eq.~(\ref{eq.Kfunc}) up to $|T|=12(eV)^{-1}$.
As already mentioned, the reduction factor due to the core-hole lifetime
is estimated as $\exp(-\Gamma_{1s}|T|)=0.000067$.
Figure \ref{fig.abs_abc} shows the calculated absorption coefficient.
The broken lines represent the quantity without taking
account of the screening by electron-hole pairs [$C(t)=1$ in 
Eq.~(\ref{eq.Kfunc})]. The curves reproduce well the curves with 
$|T|\to\infty$ shown in Fig.~\ref{fig.4pdos},
indicating that the introduction of the cutoff $|T|=12(eV)^{-1}$ causes
only minor errors.
The thick and thin solid lines represent the absorption coefficient 
for $V=6$ eV and $9$ eV.
The $K$-edge peak moves about $1.5$ eV for $V=6$ eV and about $3$ eV
for $V=9$ eV to lower-energy positions,
which may be called ``well-screened" peaks. 
In addition, other peaks appear at about an $8$ eV-higher position for $V=6$
eV and at about an $11$ eV-higher position for $V=9$ eV, in the $z$-polarization
curves, probably owing to antiscreening.
These peaks may be called ``poorly screened" ones. 
However, the poorly screened peaks are weak with broad widths 
and nearly merged into the background.
Therefore, the screening effect may be, on the whole, taken into account
by a renormalization of the core-level energy.

\subsection{RIXS intensity}

\begin{figure}
\includegraphics[width=8.0cm]{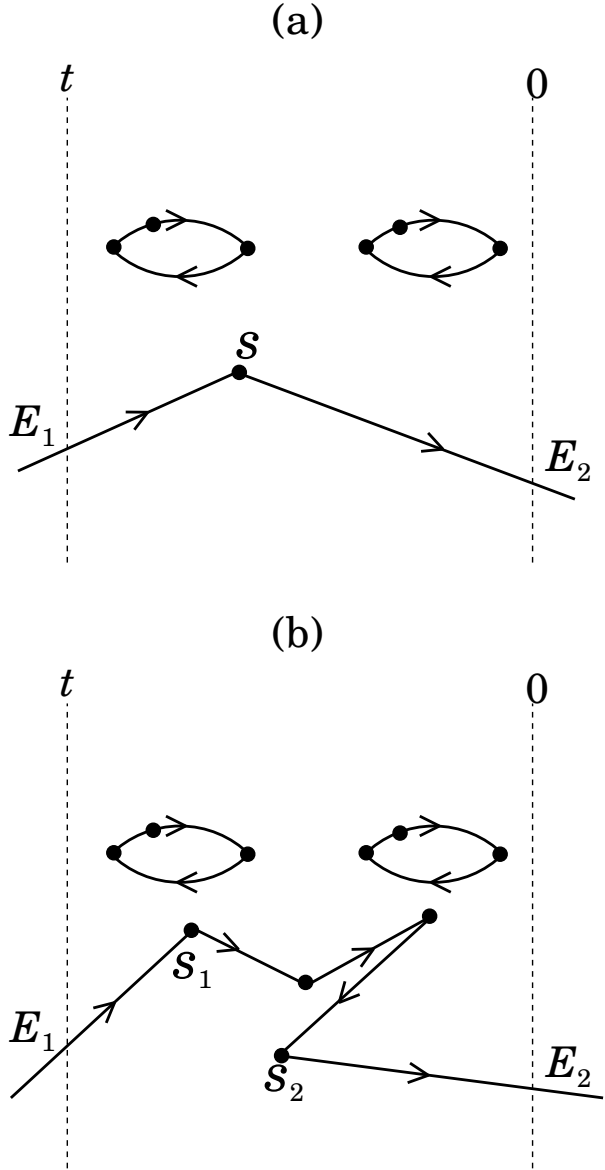}%
\caption{\label{fig.diagram_rixs_bborn}
Multiple scattering process for RIXS;(a)the Born term accompanying 
electron-hole pairs, and (b)a multiple scattering term.
The solid lines represent the one-electron Green's function $G^{\sigma}_0$
and solid circles represent the core-hole potential causing scattering.
}
\end{figure}

Confining ourselves within the HFA, we study the multiple scattering effect 
on the RIXS spectra in this subsection.
Figure \ref{fig.diagram_rixs_bborn} shows the effect of the core-hole
potential. Both the Born and the multiple scattering terms accompany 
the excitation of electron-hole pairs.
The lines with $E_1$ and $E_2$ denote the factors ${\rm e}^{-iE_1s_1}$
and ${\rm e}^{iE_2s_2}$ coming from the Keldysh-type Green's function,
$Y^{+-(0)}_{\mu'\sigma',\mu\sigma}(q)$,
which combines the outward and backward time legs. 
The product of Green's functions of the $4p$ electron and the core hole,
which is hidden in the background, gives a factor
$\exp[i(\epsilon_{4p}^{\eta}({\bf k})-\epsilon_{1s}-i\Gamma_{1s}-\omega_i)t]$.
The electron-hole pairs give an extra factor ${\rm e}^{C(t)}$, 
which is the same as in the absorption 
coefficient.  Combining these factors, we obtain
\begin{eqnarray}
 && L^{\eta}(\omega_i;E_1,E_2) =
    \sum_{\bf k}\int_{-\infty}^{0}
       \exp[i(\epsilon_{4p}^{\eta}({\bf k})-\epsilon_{1s}
             -i\Gamma_{1s}-\omega_i)t]{\rm e}^{C(t)} {\rm d}t\nonumber\\
  &\times&\left\{V\int_{t}^{0}{\rm e}^{i(E_2-E_1)s}{\rm d}s
  -V^2\int_{t}^{0}{\rm d}s_2\int_{t}^{0}
         {\rm e}^{iE_2s_2}\varphi^{\sigma}(s_2,s_1)
	 {\rm e}^{-iE_1 s_1}{\rm d}s_1 
  \right\}\\
  &=&\int_{-\infty}^{0} \left[K^{\eta}(\omega_i;-\infty)-
                              K^{\eta}(\omega_i;t)\right]{\rm d}t \nonumber\\
  &\times& \left\{V{\rm e}^{i(E_2-E_1)t}-V^2\int_{t}^{0}
     \left[{\rm e}^{iE_2 s}\varphi^{\sigma}(s,t){\rm e}^{-iE_1 t}
    +{\rm e}^{iE_2 t}\varphi^{\sigma}(t,s){\rm e}^{-iE_1 s}\right]{\rm d}s
    \right\},
\end{eqnarray}
where $K^{\eta}(\omega_i;t)$ is defined by Eq.~(\ref{eq.Kfunc}).
The multiple scattering effect is included in $\varphi^{\sigma}(s_2,s_1)$.
We finally obtain the formula of the RIXS intensity with replacing
$L_B^{\eta}(\omega_i;\omega)$ by 
$L^{\eta}(\omega_i;E_j({\bf k}),E_{j'}([{\bf k+q}]))$
in Eq.~(\ref{eq.general}).

\begin{figure}
\includegraphics[width=8.0cm]{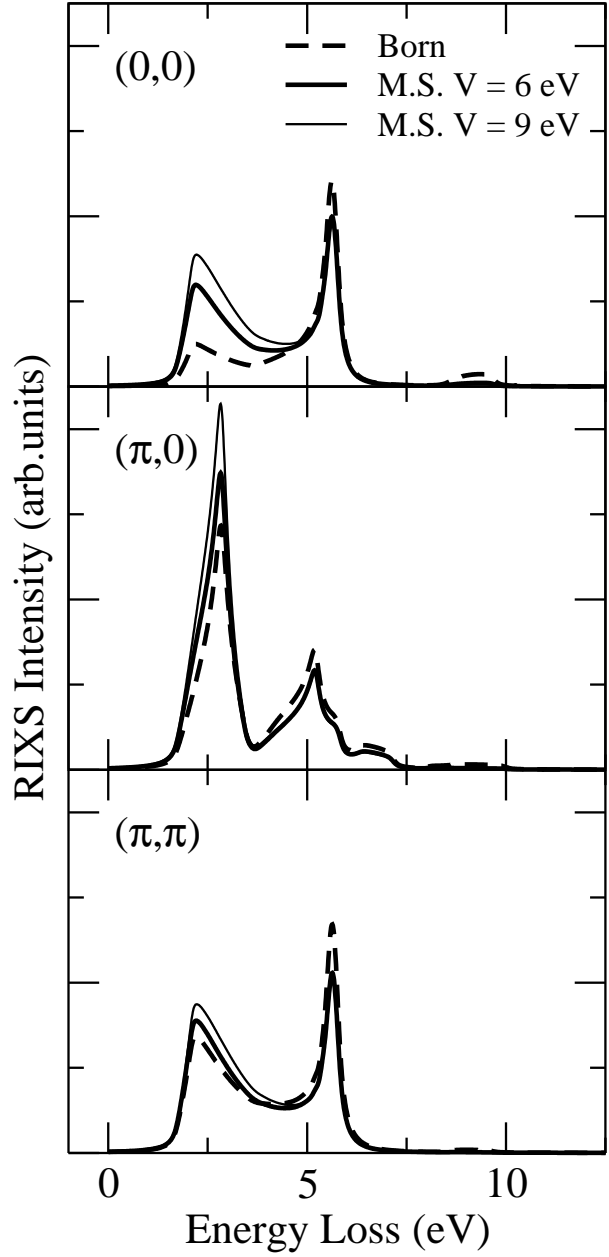}%
\caption{\label{fig.spec_multi}
RIXS spectra as a function of energy loss for momentum transfer $(0,0)$,
$(\pi,0)$, and $(\pi,\pi)$, with the inclusion of the multiple scattering 
effect. The RPA correction is not included.
The thick and thin lines represent the intensities divided by $V^2$ 
for $V=6$ and $9$ eV, respectively. 
The incident photon is assumed to have the $z$ polarization and the energy 
corresponding to the peak in the absorption spectra shown 
in Fig.~\ref{fig.abs_abc}. 
The broken lines represent the intensities divided by $V^2$
within the Born approximation, where the incident energy is set to be the 
unscreened $K$-edge peak. Dividing the intensities by $V^2$ makes the results
of the Born approximation independent of the value of $V$.
}
\end{figure}

On the basis of this formula, we calculate the RIXS spectra.
We set the core-hole potential to be $V=6$ and $9$ eV, 
and tune the incident photon energies to correspond to the peak in the
absorption coefficient  shown in Fig.~\ref{fig.abs_abc},
which are shifted from the unscreened positions.
Figure \ref{fig.spec_multi} shows the calculated spectra divided by $V^2$.
Dividing by $V^2$ makes the spectra of the Born approximation
independent of the value of $V$.

The 2 eV peak is enhanced and the 5 eV peak is suppressed by the multiple 
scattering. This tendency becomes stronger with increasing values of $V$.
This may be related to the fact that the incident photon energy is tuned to
the well-screened peak, where the excitation from the charge-transfer band
to the upper Hubbard band is most effective for screening. 
Nevertheless, the correction remains minor on the RIXS spectral shape.
By comparison with the result in the previous section, the RPA correction
is found more important.

\section{\label{sect.6}Concluding Remarks}

We have developed a theory of the RIXS on the basis of 
the Keldysh Green's function formalism.
In this formula, we have described the electronic structure 
by the $d$-$p$ model within the HFA, and have treated the electron correlation 
within the RPA. The RIXS spectra can be interpreted as
a band-to-band transition reinforced by the RPA. 
We have analyzed the RIXS spectra of La$_2$CuO$_4$ using the realistic 
$4p$ DOS, in good agreement with the experiment.

One of the important approximations in our theory was the use of
the Born approximation to treat the core-hole potential in the intermediate 
state. The multiple scattering effect may not be small, since the core-hole
potential is by no means small. For evaluating the multiple-scattering effect,
we have invented a numerical method along the line of Nozi\`eres and 
De Dominicis. In the absorption coefficient,
we have found that the $K$-edge peak moves to a lower-energy region
due to screening (well-screened peak), and that another peak appears
at a higher-energy region due to antiscreening (poorly screened peak).
However, the latter peak is weak with a broad width and nearly merged into
the background. On the RIXS spectra, we have found that 
the multiple scattering modifies the spectral shape slightly.
These findings suggest that the multiple scattering effect could be mainly
included into a renormalization of the core-level energy, and partly 
justify the use of the Born approximation. 
This paper is the first attempt to evaluate multiple scattering effects.
The above conclusion is important, since the Born approximation is
employed without examining the validity in several attempts of the RIXS
analysis.
The Born approximation is expected to work well in other insulating 
transition-metal compounds, such as NiO, LaMnO$_3$, etc.
The present formula seems promising to analyze the RIXS spectra for
such systems, because it can deal with multiband tight-binding models 
in three dimensions. 

The HFA and RPA are known to work well in the presence of the AF long-range
order. Upon doping, the AF order is easily destroyed in cuprates.
Recently the doping effects on the RIXS spectra have been observed in cuprates.
\cite{Kim04-2,Ishii05-1,Ishii05-2}
In such a situation, the HFA no longer provides a good starting point,
and we have to go beyond the HFA-RPA scheme. A study along this line is now
under progress.

\begin{acknowledgments}
This work was partially supported by a Grant-in-Aid for Scientific Research 
from the Ministry of Education, Culture, Sports, Science, and Technology, 
Japan.

\end{acknowledgments}


\bibliography{revise1}

\end{document}